\documentclass[12pt ]{article}
\usepackage{amsmath}
\usepackage{graphicx}
\usepackage{hyperref}
\usepackage{lmodern}
\usepackage{amssymb}
\usepackage{booktabs}
\usepackage{tikz}
\usetikzlibrary{patterns}
\usepackage{pdfpages}

\renewcommand{\theequation}
{\arabic{section}.\arabic{equation}}

\makeatletter
\def\eqnarray{ \stepcounter{equation} \let\@currentlabel=\theequation
 \global\@eqnswtrue
 \global\@eqcnt\z@
 \tabskip\@centering
 \let\\=\@eqncr
 $$\halign to \displaywidth\bgroup\@eqnsel\hskip\@centering
 $\displaystyle\tabskip\z@{##}$&\global\@eqcnt\@ne
 \hfil$\displaystyle{{}##{}}$\hfil
 &\global\@eqcnt\tw@$\displaystyle\tabskip\z@{##}$\hfil
 \tabskip\@centering&\llap{##}\tabskip\z@\cr}
\makeatother

\newcommand{\be}{\begin{equation}}
\newcommand{\ee}{\end{equation}}

\newcommand{\beqa}{\begin{eqnarray}}
\newcommand{\eeqa}{\end{eqnarray}}
\newcommand{\nn}{\nonumber}

\def\CA {{\cal A}}

\def\CP {{\cal P}}

\begin{document}

\setlength{\baselineskip}{7mm}
\begin{titlepage}
 
\begin{flushright} 
 {\tt NRCPS-HE-73-2025} 
\end{flushright}

\begin{center}
{\Large ~\\{\it    Schwinger's Non-Commutative Coordinates\\ 
and\\
Duality Between Helicity and Dirac Quantisation Conditions \\

 }

}

\vspace{2cm}

{\sl  George Savvidy$^{2}$

\centerline{${}^2$ \sl Institute of Nuclear and Particle Physics}
\centerline{${}$ \sl Demokritos National Research Center, Ag. Paraskevi,  Athens, Greece}

}
 
\end{center}
\vspace{2cm}

\centerline{{\bf Abstract}}

 The helicity operator of massless particles has only two polarisations like it takes place for photons and gravitons.  For them not all of the 2s+1 spin magnetic quantum states exist, with two exceptions, and the spin operator ceases to be defined properly and consistently. The problem was solved by Schwinger, who introduced non-commutative space coordinates that completely eliminate the spin operator and ensure that only helicity operator appears explicitly.    We further investigate the violation of the associativity relation of the momentum translation operator that emerges due to the failure of the corresponding Jacobi identity. The associativity relation is broken by a phase factor which satisfies a 3-cocycle relation.  The associativity is restored when a 3-cocycle is an integer number, and leads to the quantisation of massless particle's helicity. We discuss the correspondence (duality) between the helicity and the Dirac quantisation conditions. The relation for the minimal space cell volume, similar to the minimal phase-space cell of Heisenberg is suggested.

  \vspace{12pt}

\noindent

\end{titlepage}

\pagestyle{plain}

\section{\it Introduction }
  The helicity operator of massless particles has only two polarisations like it takes place for photons and gravitons.  For them not all of the 2s+1 spin magnetic quantum states exist, with two exceptions, and the spin operator ceases to be defined properly and consistently. To abandon a spin operator and to ensure  that only helicity operator $\lambda$ appears explicitly in the massless representations 
Schwinger   \cite{Schwinger} introduced  the non-commutative space coordinates $ \hat{\vec{R}}$   
\be 
[\hat{R}_n ,  \hat{R}_m] =-i \hbar \lambda \epsilon_{nmk} {P_k \over P^3}, \nn
\ee
that completely eliminate a spin operator.  The corresponding uncertainty relations 
\be
\triangle \hat{R}_n~ \triangle \hat{R}_m \geq \hbar ~{\vert \lambda \vert  \over 2} ~ \Big\vert   \Big\langle {P_k \over P^3}  \Big\rangle \Big\vert, ~~~~~~~~ n \neq m \neq k, \nn
\ee
express the fact that the average wavelength of a massless particle sets the scale of the coordinate uncertainty.  Here $\triangle \hat{R}_n^2 = \langle~ ( \hat{R}_n  - \langle  \hat{R}_n \rangle)^2 ~\rangle $ is the mean square of the deviation of $\hat{R}_n$ from its mean value $\langle  \hat{R}_n \rangle$ and  $\langle {P_k \over P^3} \rangle$ is the mean value of the momentum operator.  For a momentum state with some degree of directionality along a given axis  one can obtain that 
\be\label{uncertanty}
(\triangle \hat{\textbf{R}}  )^2   \geq    \hbar  ~\vert \lambda \vert   ~  \Big\langle  { 1 \over P^2   }  \Big\rangle,  \nn 
\ee
where $ {2 \pi \hbar \over P} =  \Lambda $ is the wavelength of the massless particle and indicate that the wavelength sets the scale of coordinate uncertainty \cite{ Schwinger, Poincare, Wigner:1939cj} .

We further investigate the violation of the associativity relation of the momentum translation operator $
U(\vec{b}) = e^{-{i\over \hbar} \vec{b} \hat{\vec{R}}} 
$ emerging due to the failure of the infinitesimal generators to satisfy the Jacobi identity, which is obstructed by the zero-momentum of a massless particle: 
\be 
~[[\hat{R}_1 ,  \hat{R}_2]  ,\hat{R}_3]+[[\hat{R}_3 ,  \hat{R}_1]  ,\hat{R}_2]+ [[\hat{R}_2 ,  \hat{R}_3]  ,\hat{R}_1] = - 4 \pi \lambda   \hbar^2 \delta^{(3)}(\vec{P}). \nn
\ee
We investigated  the associativity relation 
 \be\label{assosi}
U(\vec{b}_1) ~( U(\vec{b}_2) U(\vec{b}_3) ) = e^{i\Phi/\hbar} ~(U(\vec{b}_1) ( U(\vec{b}_2)) ~U(\vec{b}_3) 
 \ee
and obtained that the phase $\Phi$ is the total flux of the momentum ${\vec{P} \over P^3} $ through the surface of tetrahedron constructed by the three momentum vectors  $(\vec{b}_1, \vec{b}_2,\vec{b}_3)$ (see Fig.\ref{fig1})
\be
\Phi = \lambda \oint {\vec{P} \over P^3} d \vec{S}= \lambda \int \Big( \vec{\nabla}\cdot {\vec{P} \over P^3} \Big)  ~d^3 \vec{P}=
\lambda \int  4 \pi \delta^{(3)}(\vec{P}) ~d^3 \vec{P} = 4 \pi \lambda .\nn
\ee
The associativity relation (\ref{assosi}), which is necessary for conventional quantum mechanics on the operators on Hilbert space, is broken by a phase factor $\Phi$ which satisfies a 3-cocycle relation.   
The quantum mechanical associativity (\ref{assosi}) will be restored if the phase $\Phi$ is proportional to $2 \pi n$ 
\be\label{quanhell}
{\Phi \over \hbar } = {4 \pi \lambda\over \hbar }   = 2 \pi n, ~~~~~\lambda = {\hbar \over 2} n,~~~n = \pm 1, \pm 2, ..... 
\ee
{\it Thus the quantum mechanical consistency requires the quantisation of particle's helicity.} 
 
We analyse the correspondence between the helicity quantisation condition (\ref{quanhell}) and the Dirac quantisation condition for a charged particle moving in a background monopole field.   It is well known that the canonical momentum $\vec{p}$ of a particle of mass $m$ and charge $e$ moving in a magnetic field of a monopole of charge $g_m$  has  the commutation relations of the form  
\beqa
~[p_n, p_m]= i \hbar {e\over c} \epsilon_{nmk} H_k,\nn
\eeqa
where the magnetic field of a monopole is 
$
\vec{H} =  g_m  \vec{r} / r^3.
$
One can verify that the Jacobi identity failed at the origin $\vec{r}=0$ for the triple momentum commutator and has the following form:
\be
~[[p_1,p_2],p_3] + [[p_2,p_3],p_1] +[[p_3,p_1],p_2] = 
- 4 \pi \hbar^2 {e g_m\over  c} ~\delta^3(\vec{r}).\nn
\ee 
 The associativity relation for the translation operator 
$
U(\vec{a}) = e^{{i\over \hbar} \vec{a} \vec{p}}
$
is broken here as well by a phase factor $\Phi_m$  which is the total flux of the vector field  ${\vec{r} \over r^3} $ through the surface of tetrahedron constructed by the space-like vectors  $(\vec{a}_1, \vec{a}_2,\vec{a}_3)$ \cite{Jackiw:1984rd}
\be
\Phi_m ={e g_m \over c} \oint {\vec{r} \over r^3} d \vec{S}= {e g_m \over c}\int \Big(\vec{\nabla}\cdot {\vec{r} \over r^3}\Big)  ~d^3 \vec{r}=
{e g_m \over c} \int  4 \pi \delta^{(3)}(\vec{r}) ~d^3 \vec{r} = 4 \pi {e g_m \over c} \nn
\ee
The quantum mechanical associativity will be restored if this phase is proportional to $2 \pi n$ 
\be
{\Phi_m \over \hbar } = 4 \pi  {e g_m \over \hbar c}   = 2 \pi n, ~~~~~{e g_m \over c}  = {\hbar \over 2} n,~~~n = \pm 1, \pm 2, .....
\ee
{\it The removal of the 3-cocycle, which is necessary for conventional quantum mechanics on the operators on Hilbert space, limits magnetic sources to quantised Dirac monopoles.}  These relations are in a well defined correspondence  with the helicity quantisation condition (\ref{quanhell}) of the massless particles considered above  if one consider a duality map 
\be
\hat{\vec{R}} \longleftrightarrow  \vec{p},~~~~\vec{P}  \longleftrightarrow - \vec{r}, ~~~~~~~~\lambda  \longleftrightarrow {e g_m\over  c}. 
\ee
In the subsequent four sections we discuss the representations of the Poincar\'e algebra  \cite{ Schwinger, Poincare, Wigner:1939cj} and in the sixth section we introduce the Schwinger non-commutative coordinates for the massless representations.   In the seventh section we analyse the associativity relation which is broken by a 3-cocycle and derive the helicity quantisation condition\footnote{The discussion of group cocycles can be found in  \cite{Zumino:1983ew, Mickelsson:1983xi, Faddeev:1984jp, Zumino:1983rz, Jackiw:1984rd, Arakelian:1988gm, Savvidy:2010bk, Antoniadis:2012ep, Antoniadis:2013jja,  Konitopoulos:2014owa}.}.  In the eighth section we discuss the correspondence (duality) between the helicity quantisation and the Dirac quantisation conditions. In the ninth section the consequences of non-commutativity of photon and graviton space coordinates are discussed and the relation for the minimal space cell volume, similar to the minimal phase-space cell of Heisenberg is suggested. The high-spin extension of Poincar\'e algebra and its massless representations are considered in the tenth section. 
 
\section{\it Poincar\'e algebra} 
 
 The fundamental property of the spacetime is formulated as the invariance of all physical systems with respect to the corresponding  infinitesimal coordinate transformations defined in the following way\footnote{$x^{\mu} =(c t, \vec{r})$, $x^0 =- x_0 =c t $, $x^i = x_i = r_i$. }:
\be
\delta x^{\mu} = \delta \epsilon^{\mu} + \delta \omega^{\mu\nu} x_{\mu},~~\delta \omega^{\mu\nu} = -\delta \omega^{\nu\mu},
\ee
where $ \delta \epsilon^{\mu}$  is a spacetime translation and $\delta \omega^{\mu\nu}$  is a  four-dimensional rotation.  The six independent parameters of this four-dimensional rotation are related to $\delta \vec{\omega}$ and $\delta \vec{v}$  by
\be
\delta \omega_{ij}  =\epsilon_{ijk} \delta \omega_k,~~~~\delta \omega_{0i}  = {1\over c } \delta v_i ,
\ee
where $\vec{\omega}$ is the angular velocity and $\vec{v}$ is the velocity. The infinitesimal unitary transformation, $U = 1 + i G$ , that is induced by  an infinitesimal coordinate transformation is given by the expression  
\be
U = 1 +i {1\over \hbar }(P^{\mu} \delta \epsilon_{\mu} + {1\over 2} M^{\mu\nu}\delta \omega_{\mu\nu}), 
\ee
where the space components of the Poincar\'e  generators $P^{\mu} $ and $M^{\mu\nu}$ are 
\be
cP^0 = H + M c^2,~~~~~~    P_i,~~~~~{1\over c} M^{0i}= N_i, ~~~~~M_{ij}=\epsilon_{ijk} J_k.
\ee
The generators $P_i,$ and $J_k,$ are   the linear and angular momentum operators, while H is the energy, or the Hamiltonian operator, and $N_i$ is the boost operator, in total ten operators.   The full set of commutators for the generators comprise the Poincar\'e algebra  \cite{ Schwinger, Poincare, Wigner:1939cj}:
\beqa\label{gaugePoincare}
~&&[P_{\mu},~P_{\nu}]=0,\nn\\
~&&[P_{\mu},~ M_{\kappa\lambda},~] = i \hbar (\eta_{\mu\lambda }~P_{\kappa}
- \eta_{\mu \kappa}~P_{\lambda}) ,\nn\\
~&&[M_{\mu \nu}, ~ M_{\kappa\lambda }] = i \hbar(\eta_{\mu \kappa}~M_{\nu \lambda}
-\eta_{\nu \kappa}~M_{\mu \lambda} +
\eta_{\nu \lambda}~M_{\mu \kappa}  -
\eta_{\mu \lambda}~M_{\nu \kappa} ),
\eeqa
where 
\be
\eta_{00}=-1,~~~ \eta_{0i}=0, ~~~\eta_{ij}=\delta_{ij}. \nn
\ee
The commutators can also be presented in the following  equivalent form:
\beqa
~[P^{\nu}, {1\over 2} M^{\kappa\lambda} \delta \omega_{\kappa\lambda}]=i \hbar \delta \omega^{\mu\nu} P_{\mu}, ~~~~~~~~~~~~~~~~~~~~~~~~~~~~&& ~[P^{\nu},   P^{\lambda} \delta \epsilon_{\lambda}]=0 \\
~[M^{\mu\nu}, {1\over 2} M^{\kappa\lambda} \delta \omega_{\kappa\lambda}]= i \hbar (\delta \omega^{\lambda\mu} M^{~~\nu}_{\lambda}+ \delta \omega^{\lambda\nu} M^{~~\mu}_{\lambda}),~~~~ && ~[M^{\mu\nu}, P^{\lambda} \delta \epsilon_{\lambda}]=i \hbar(\delta \epsilon^{\lambda} P^{\nu} - i \delta \epsilon^{ \nu} P^{\mu} )\nn,
\eeqa
indicating the response of vectors and tensors to infinitesimal Lorentz rotations,  here ${1\over 2} M^{\kappa\lambda} \delta \omega_{\kappa\lambda}$ corresponds  to the three-dimensional rotations and boosts.  The response of these operators to the translational $P^{\lambda} \delta \epsilon_{\lambda}$ is given as well.  In terms of a three-dimensional vector the algebra will take the following form: 
\beqa\label{gaugePoincare3d}
~&&[P_{n},~P_{m}]=0,  ~~~~~~~~~~~~~~~~~~~[J_{n}, J_{m}]=i \hbar\epsilon_{nmk} J_k, \label{JJ} \\
~&&[P_{n},~J_{m}]=i \hbar \epsilon_{nmk} P_k,  ~~~~~~~~~[N_{n},~N_{m}]= -i \hbar {1\over c^2}   J_{nm},\label{spin} \\
~&&[N_{n},~J_{m}]=i \hbar \epsilon_{nmk} N_k, ~~~~~~~~[P_{n},~N_{m}]=i \hbar \delta_{nm} {P_0 \over c},\label{NN} \\
 ~&&~~~~~~~~~~~~~~~~~~~[H,~P_{n}]=0,\nn\\
~&&~~~~~~~~~~~~~~~~~~~[H,~J_{n}]=0,\nn\\
~&&~~~~~~~~~~~~~~~~~~~[H,~N_{n}]=i \hbar P_n, \label{JNS}
\eeqa
When  ${P_0 \over c} \ll 1$, the algebra reduces to the Galilean  algebra.

\section{\it Casimir invariants  of Poincar\'e algebra   }

The  fully invariant vacuum state $ \vert 0 \rangle$ is defined as 
\be
P^{\mu}  \vert 0 \rangle =0,~~~~M^{\mu\nu}  \vert 0 \rangle =0.
\ee
The excited states will have a positive value of $P_0 > 0$. Let us consider the first invariant quantity 
\be\label{timelike}
- P^{\mu}P_{\mu} = (P^0)^2 - \vec{P}^2  = M^2 c^2, 
\ee
which  can be strictly positive $M^2 >0$, and therefore i) $P^0 =+ (\vec{P}^2  +M^2 c^2)^{1/2} >0$ or ii) equal to zero $M^2 =0$ and then   $P^0 =+ \vert  \vec{P}\vert >0 $. The second invariant can  be constructed by using the Pauli-Lubanski pseudo-vector 
\be\label{paulilubanski}
W^{\mu} ={1\over 2} \epsilon^{\mu\nu\lambda\rho} M_{\nu\lambda} P_{\rho}.~~~~
\ee
It is a translationally invariant  pseudo-vector 
\be
~[W^{\mu},  P^{\nu} \delta \epsilon_{\nu}, ]= \hbar \epsilon^{\mu\nu\lambda\rho} \delta \epsilon_{\lambda} P_{\rho} P_{\nu}=0,
\ee
and it has the following properties:
\be\label{noncommutative}
[P^{\mu},~W^{\nu}]=0~,  ~~~[W^{\mu},~W^{\nu}]=
i \hbar \epsilon^{\mu\nu\lambda\rho}~ W_{\lambda} P_{\rho}~,~~~~~~~P_{\mu} W^{\mu}=0.
\ee
That is,  the Pauli-Lubanski pseudo-vector is translationally invariant, non-commutative pseudo-vector,  and because it is orthogonal to the time-like vector $P^{\mu}$ (\ref{timelike}) it is a space-like pseudo-vector:
\be
~~W^{\mu}W_{\mu} =\varrho^2 \geq 0.
\ee
The above two Casimir invariants can be used to characterise the unitary irreducible  representations of the Poincar\'e algebra.

\section{\it Representations of Poincar\'e algebra  without spin }

We are interested in describing the representations of the Poincar\'e algebra in terms of operators acting in an appropriate Hilbert space. It follows from (\ref{spin}) and (\ref{NN}) that the angular momentum and boost  operators transform under the coordinate translations  as follows:
\beqa
\delta_{\epsilon}\vec{J} &=&~{1\over i \hbar }[\vec{J}, ~\vec{P}\cdot \delta \vec{\epsilon}] = \delta\vec{\epsilon} \times \vec{P}\\
\delta_{\epsilon} \vec{N} &=&~{1\over i \hbar}[\vec{N} , ~\vec{P}\cdot \delta \vec{\epsilon}]  = - \delta \vec{\epsilon}   ~{P^{0}  \over c}.
\eeqa
The Poincar\'e algebra has the operators $(P^0,\vec{P})$,  the boost $\vec{N}$ and angular momentum $  \vec{J}$ operators, but there are no explicit coordinate operators in the algebra. The response of the angular momentum to translation is in accordance with the nature of angular momentum 
and indicates the existence of a position vector operator $\vec{R}$:
\beqa
\delta_{\epsilon}\vec{R} &=&~{1\over i \hbar}[\vec{R}, ~\vec{P}\cdot \delta \vec{\epsilon}] = \delta\vec{\epsilon},
\eeqa
thus having the Heisenberg commutator with momentum operators $\vec {P}$:
\be
[ R_i, P_k] = i \hbar \delta_{ik}.
\ee
The next step is to construct the representation of the $\vec{J}$ and $\vec{N}$ operators in terms of coordinate and momentum operators.  The angular momentum and boost operators can be defined as 
\be\label{representaion0}
\vec{J} = \vec{R} \times \vec{P},~~~~\vec{N}= \vec{P} x^0 - {1\over c} \{P^0, \vec{R}\}, 
\ee  
where the product $P^0 \vec{R} $ is symmetrised  because these operators are not commuting:
\be
{1\over i \hbar }[\vec{R}, P^0] =   {\partial P^0 \over \partial \vec{P}} =  { \vec{P} \over  P^0}. 
\ee
Importantly, only when  the coordinate operators are commuting operators:
\be\label{coordcomm}
[R_i , R_j] =0,
\ee
then all constructed operators are correctly transforming with respect to the rotations.  The most characteristic  commutator to be checked   is (\ref{spin}). One can derive that
\be
{i\over \hbar}[\{P^0 ,R_n\}, \{P^0, R_m\}]= R_n P_m - R_m P_n
\ee
and then get convinced that 
\be\label{boost}
~ i \vec{N} \times \vec{N}= i (\vec{P} x^0 - {1\over c} \{P^0, \vec{R}\})\times (\vec{P} x^0 - {1\over c} \{P^0 ,\vec{R}\})= {\hbar \over c^2} \vec{R}\times \vec{P}=  {\hbar \over c^2} \vec{J}.
\ee
The representation (\ref{representaion0}) fulfils all commutation relations of the Poincar\'e algebra (\ref{gaugePoincare3d}-\ref{JNS}) and completes the construction of the representation of the Poincar\'e algebra for particles without spin,  the  internal angular momentum $\vec{S}$. 
 
 \section{\it  Massive representation with spin}

Let us now describe the representation that contains the internal angular momentum, the spin of particles,  by adding a new term to the angular momentum $\vec{J}$ (\ref{representaion0}):
\be
\vec{J} = \vec{R} \times \vec{P} ~~~~\rightarrow ~~~~~ \vec{J} = \vec{R} \times \vec{P} +\vec{S}.
\ee
To separate these two terms in the total angular momentum $\vec{J}$ one should request that the spin operator commutes with the coordinate and momentum operators:
\be
[S_n, R_m]= [S_n,P_m]= 0,
\ee
and to keep intact  the commutator of angular momentum operators (\ref{JJ}) one should have 
\be
[S_{i},~S_{l}]=i \hbar \epsilon_{ilk} S_k
\ee
with all of the $2s+1$ spin magnetic quantum number states.  As soon as the angular momentum operator changes, the boost operator $\vec{N}$ should be redefined so that the commutator (\ref{spin}), (\ref{boost}) remains  intact:
\be
i{c^2 \over \hbar} \vec{N} \times \vec{N} = \vec{J} = \vec{R} \times \vec{P} +\vec{S}.
\ee
One can add the term  $f(P^0) ~ \vec{S} \times \vec{P}$  with an unknown function $f(P^0)$ and then find out that $f(P^0) = {1\over P^0 + M c}$, thus  the modified  boost operator $\vec{N}$ should have the following form:
\be
\vec{N}= \vec{P} x^0 -{1\over c}\{P^0 \vec{R}\}  + {1\over c^2} {1\over P^0 + M c} \vec{S} \times \vec{P}.
\ee
In summary, the massive representation that contains the  spin operator $\vec{S}$ has the following form:
\beqa 
 \vec{J} &=& \vec{R} \times \vec{P} +\vec{S}, \label{theJ}\\
  \vec{N}&=& \vec{P} x^0 - {1\over c}\{P^0 \vec{R}\}  + {1\over c} {1\over P^0 + M c} \vec{S} \times \vec{P}. \label{theN}
\eeqa
The representation (\ref{theJ} - \ref{theN}) fulfils all commutation relations of the Poincar\'e algebra (\ref{gaugePoincare3d}-\ref{JNS}) and completes the construction of the representation of the Poincar\'e algebra for particles with spin in terms of  the coordinate,  momentum and spin operators.

It is possible to invert these formulas and express the coordinate and spin operators in terms of the momentum, angular momentum, and boost operators. By calculating the products $\vec{J} \cdot \vec{P}$, ~$(\vec{J} \times \vec{P})$, ~ $\vec{N} \cdot \vec{P}$ and $(\vec{N} \times \vec{P})$ one can represent the coordinate and spin operators in the following form ($x^0=0$):
\beqa\label{iverseforms}
M \vec{R} &=& -\vec{N} + {1\over P^0 (P^0 +Mc )} \vec{P} (\vec{P} \cdot \vec{N}) + {1\over P^0 +M c } \vec{J} \times \vec{P}, \nn\\
M \vec{S} &=& {1\over c} P^0 \vec{J} - {1\over c}{1\over P^0 +M c} \vec{P} (\vec{P} \cdot \vec{J}) +  \vec{N} \times \vec{P}.
\eeqa
The components of  the Pauli-Lubanski vector (\ref{paulilubanski}) in this representation are:
\beqa\label{componentsPW}
W^0&=& \vec{P} \cdot \vec{J}=\vec{P} \cdot \vec{S},~~~\nn\\
\vec{W} &=& P^0 \vec{J} -  c\vec{P}  \times \vec{N} = c M \vec{S} +{(\vec{P} \cdot \vec{S})\over P^0 +M c } \vec{P} . 
\eeqa 
It follows that the second invariant of the Poincar\'e algebra is proportional to the square of the spin of a particle:
\be\label{totspin}
W^2 = c^2 M^2 \vec{S}^2,
\ee
and that $\vec{S}^2$ is a Lorentz invariant quantity.  In summary, the representation is characterised by two invariants, the mass of the particles $-P^2 =  M^2 c^2$ (\ref{timelike}) and their spin (\ref{totspin}). This discussion refers to a strictly massive case $M^2 >0$.

\section{\it  Massless helicity states  and non-commuting coordinates }

Our main interest is to consider the massless particles, such as photons, gravitons, and high helicity states (\ref{tensorspectrum}).  We will consider the limit $M^2 \rightarrow 0$ of the massive representation described above when $\vec{S}^2$ is kept fixed. In this limit  when $M^2= - P^{\mu} P_{\mu} =0$, from (\ref{componentsPW}) we will have 
\be
W^0= \vec{P} \cdot \vec{S},~~~\vec{W} =   {(\vec{P} \cdot \vec{S})\over P} \vec{P} ,~~~
 W^2 = 0,  
\ee
The above relations can be written in the following form:
\be
W^{\mu}= \lambda P^{\mu},
\ee
where $\lambda$ is the Lorentz invariant helicity operator, which has the following form:
\be
\lambda = {\vec{P} \cdot \vec{S} \over P}.
\ee
It is equal to the components of the spin along the direction of the motion, and as far as it is Lorentz invariant, the system exhibits only two values of helicity $\pm s$. {\it This means that not all of the $2s+1$ spin magnetic quantum number states exist in the massless limit.} 

Thus in the limit $M^2 \rightarrow 0$  the spin operator ceased to be defined, with two exceptions,  and one should {\it introduce new variables} for this circumstance. In order to eliminate the operator $\vec{S}$ from a massless representation Schwinger suggested that the new coordinates can be defined in the following form \cite{Schwinger} :
\be\label{newcoor}
\hat{\vec{R}}= \vec{R} - {\vec{S} \times \vec{P} \over P^2}.
\ee
In that case the commutation relation of the coordinate $\hat{\vec{R}} $ and the momentum operator $ \vec{P}$ remains intact:
\be\label{heizenberg2}
~[P_n , P_m]=0, ~~~~[\hat{R}_n , P_m]= i \hbar \delta_{nm}, 
\ee
but the commutation relation (\ref{coordcomm})  between the coordinates will change and take  the following form:
\be\label{masslesscood}
[\hat{R}_n ,  \hat{R}_m] =-i \hbar \lambda \epsilon_{nmk} {P_k \over P^3}, 
\ee
which implies that the Jacobi identity is obstructed by zero-momentum particles and will takes the following form:
\be\label{jacobi}
~[[\hat{R}_1 ,  \hat{R}_2]  ,\hat{R}_3]+[[\hat{R}_3 ,  \hat{R}_1]  ,\hat{R}_2]+ [[\hat{R}_2 ,  \hat{R}_3]  ,\hat{R}_1] =  \lambda \hbar^2  \triangle_{ P} \Big({1\over P}\Big) =   - 4 \pi \lambda   \hbar^2 \delta^{(3)}(\vec{P}).
\ee
It imposes the restriction $\vec{P} \neq 0$  validating the Lorentz invariant energy  property $P^0 = \vert \vec{P} \vert >0$. The absence of certain values of spin magnetic quantum number states is now manifested by the noncommutativity of $\hat{\vec{R}}$ components.

As one can see, the new coordinates are not commuting between themselves, and this is opposite to the massive case where the coordinate operators are commuting (\ref{coordcomm}). The vector product of the new coordinates with the momentum operator is such that 
\be 
\hat{\vec{R}}  \times \vec{P} =  \vec{R} \times \vec{P} +\vec{S} -  {\vec{P} \cdot \vec{S} \over (P^0)^2} \vec{P}= \vec{J} -  \lambda  {\vec{P}  \over P  },
\ee
and we can express the angular momentum operator in terms of new coordinates and helicity operators:
\be\label{masslessmom}
\vec{J} = \hat{\vec{R}}  \times \vec{P}   + \lambda  {\vec{P}  \over P }. 
\ee
It is also true that 
\be
[\lambda,  \hat{\vec{R}}]=0.
\ee
The other commutation relations are given in the Appendix.  It follows from this equation that in terms of the new coordinate  operators the massless representation of the angular operator $\vec{J} $ can be written only in terms of the helicity operator without any reference to spin operators.  From the equation (\ref{theN}) it follows that the boost operator $\vec{N}$ will take the following form:
\be\label{masslessmom1}
\vec{N}= \vec{P}x^0 - {1\over c}   P^0 \hat{\vec{R}} .
\ee
The expressions   (\ref{masslesscood}), (\ref{masslessmom}) and (\ref{masslessmom1}) completely define the massless representation of the Poincar\'e algebra in terms of new non-commuting coordinates $\hat{\vec{R}}$, momentum $\vec{P}$,  and helicity operator $\lambda$.  If the explicit expresions of $\vec{J}$ and $\vec{N}$  are inserted in the formulas for $M \vec{R}$ and $M \vec{S}$, the expressions (\ref{iverseforms}) do vanish, as does $M \hat{\vec{R}}$.
 
The intrinsic non-locality of massless particles (\ref{masslesscood}) is described by the Heisenberg uncertainty relation:
\be\label{basicuncertanty}
\triangle \hat{R_n}~ \triangle \hat{R_m} \geq \hbar {\vert \lambda \vert  \over 2}  \Big\vert  \Big\langle {P_k \over P^3}  \Big\rangle \Big\vert, ~~~~~~~~ n \neq m \neq k,
\ee
where $\triangle \hat{R_n}^2 = \langle~ ( \hat{R_n}  - \langle  \hat{R_n} \rangle)^2 ~\rangle $ is the mean square of the deviation of $\hat{R_n}$ from its mean value $\langle  \hat{R_n} \rangle$ and  $\langle {P_k \over P^3} \rangle$ is the mean value of the operator in the state under consideration.  For a momentum state with some degree of directionality along a given axis one can obtain that 
\be\label{uncertanty}
(\triangle \hat{\textbf{R}}  )^2   \geq    \hbar ~ \vert \lambda \vert~  \Big\vert     \Big\langle  { 1 \over P^2   }  \Big\rangle    \Big\vert ,   
\ee
where $ {2 \pi \hbar \over P} =  \Lambda $ is the wavelength of the massless particle indicating that the wavelength sets the scale of coordinate uncertainty. 

The other interesting quantity would be the volume uncertainty that can be obtained by the multiplication of all the three relations from (\ref{basicuncertanty}) \footnote{The equation (\ref{volumeuncer}) was derived by calculating the product of all the three relations that follow from the main uncertainty relations (\ref{basicuncertanty}) at $k=1,2,3$.}:
\be\label{volumeuncer}
\triangle \hat{R_1}^2~\triangle \hat{R_2}^2~\triangle \hat{R_3}^2~ \geq   \hbar^3 ~{\vert \lambda \vert^3  \over 8}  ~\Big\vert  \Big\langle {P_1 \over P^3}  \Big\rangle   \Big\langle {P_2 \over P^3}  \Big\rangle  \Big\langle {P_3 \over P^3} \Big\rangle \Big\vert ,
\ee 
and if one speculates that the graviton $\lambda = 2 \hbar$ can have a wave length  always 
larger than the Planck length $P \approx \hbar/ l_{Pl}$, one can get the minimal space volume of the order 
\be\label{quantvol}
\triangle \hat{R}_1~\triangle  \hat{R}_2~\triangle   \hat{R}_3 ~ \geq  \Big(  {G \hbar \over c^3}   \Big)^{3/2}.
\ee
The space-time quantisation was considered in the earlier publications in \cite{Doplicher:1994zv,Doplicher:1994tu, Yoneya:2000bt,Lust:2010iy,Mylonas:2013jha, Bacry:1987sa}.

The uncertainty relation (\ref{volumeuncer}) can  be written in an invariant form by  considering a momentum state with some degree of directionality along a given axis, let us say, the third axis.    From the commutation relations (\ref{heizenberg2}) and (\ref{basicuncertanty})  we can obtain that      
\be\label{basicuncertanty1}
\triangle \hat{R}_1~ \triangle \hat{R}_2 \geq \hbar ~{\vert \lambda \vert  \over 2} ~ \Big\vert   \Big\langle {1 \over P^2_3}  \Big\rangle \Big\vert,~~~~~
\triangle \hat{R}_3~ \triangle  P_3 \geq  \frac{\hbar}{2}  
\ee
and therefore 
\be
\triangle \hat{R}_1^2~\triangle  \hat{R}_2^2~\triangle   \hat{R}_3^2~ \geq    \hbar^4  ~  \frac{\lambda^2}{16}    ~   \frac{1 }{ \triangle  P^2_3} ~\Big\vert  \Big\langle {1 \over P^2_3}  \Big\rangle \Big\vert^2.
\ee
Because $\triangle  P^2_3 \leq  \triangle  P^2_1+\triangle  P^2_2 +\triangle  P^2_3 \equiv  (\triangle \textbf{P} )^2 $ and 
$$
\Big\vert  \Big\langle {1 \over P^2_3}  \Big\rangle \Big\vert  \geq   \Big\vert   \Big\langle \frac{1}{ P^2_1 + P^2_2+ P^2_3}   \Big\rangle \Big\vert  \equiv  \Big\vert  \Big\langle \frac{1}{ P^2}   \Big\rangle \Big\vert 
$$
we will obtain the invariant form of the uncertainty  relation (\ref{volumeuncer}):
\be\label{volumeuncer1}
\triangle \hat{R}_1^2~\triangle  \hat{R}_2^2~\triangle   \hat{R}_3^2~ \geq    \hbar^4  ~  \frac{\lambda^2}{16}    ~   \frac{1 }{(\triangle \textbf{P} )^2} ~\Big\vert  \Big\langle \frac{1}{ P^2}   \Big\rangle \Big\vert^2.
\ee
By multiplying the  Heisenberg uncertainty relations for the momenta $P_n$ and coordinates $\hat{R}_n$  (\ref{heizenberg2}) one can obtain the alternative uncertainty  relation   
\be\label{volumeuncer2}
\triangle \hat{R}_1^2~\triangle  \hat{R}_2^2~\triangle   \hat{R}_3^2~ \geq     \frac{ \hbar^6 }{64}    ~   \Big(\frac{1 }{(\triangle \textbf{P} )^2} \Big)^3 
\ee
and  recognise that it gives a less rigorous lower bound on minimal space volume than (\ref{volumeuncer1}) because the important dependence on the wave length $~\Big\vert  \Big\langle \frac{1}{ P^2}   \Big\rangle \Big\vert^2$ present in (\ref{volumeuncer1}) is absent here.

 \section{\it Violation of associativity relation by 3-cocycle }
 
As we have seen the commutation relation of the coordinate $\hat{\vec{R}} $ and the momentum operator $ \vec{P}$ remains intact:
\be\label{heizenberg3}
~[P_n , P_m]=0, ~~~~[\hat{R}_n , P_m]= i \hbar \delta_{nm}, 
\ee
but the commutation relation (\ref{coordcomm})  between the coordinates will change and take the following form:
\be\label{masslesscood1}
[\hat{R}_n ,  \hat{R}_m] =-i \hbar \lambda \epsilon_{nmk} {P_k \over P^3}, 
\ee
which implies that the Jacobi identity is obstructed by zero-momentum,  and will takes the following form:
\be\label{jacobi1}
~[[\hat{R}_1 ,  \hat{R}_2]  ,\hat{R}_3]+[[\hat{R}_3 ,  \hat{R}_1]  ,\hat{R}_2]+ [[\hat{R}_2 ,  \hat{R}_3]  ,\hat{R}_1]  =   - 4 \pi \lambda   \hbar^2 \delta^{(3)}(\vec{P}).
\ee
We will consider now the {\it momentum translation operator} 
\be\label{uniop}
U(\vec{b}) = e^{-{i\over \hbar} \vec{b} \hat{\vec{R}}}, ~~~~~U(\vec{b}) \Psi(\vec{P})= \Psi(\vec{P}+ \vec{b})
\ee 
and would like to analyse the associativity relation
 \be\label{associ}
U(\vec{b}_1) ~( U(\vec{b}_2) U(\vec{b}_3) ) = e^{i\Phi/\hbar} ~(U(\vec{b}_1) ( U(\vec{b}_2)) ~U(\vec{b}_3). 
 \ee
Here the non-trivial phase factor $\Phi$ satisfies a 3-cocycle relation. The 3-cocycle  $\omega_3$ appears when the associativity is violated by a phase factor that enters into the associativity relation\footnote{ The discussion of group cocycles can be found in  \cite{Zumino:1983ew, Mickelsson:1983xi, Faddeev:1984jp, Zumino:1983rz, Jackiw:1984rd, Arakelian:1988gm, Savvidy:2010bk, Antoniadis:2012ep, Antoniadis:2013jja,  Konitopoulos:2014owa, Izaurieta:2017awb, Rubio:2018itx, Song:2023rra}.}
\be
U(g_1) ~( U(g_2) U(g_3) ) = e^{2 \pi i \ \omega_3(x;g_1,g_2,g_3)} ~(U(g_1) ( U(g_2)) ~U(g_3),
\ee
where the group operators $U(g)$  implement the group action on the functions $f(x)$.   A consistency condition on $\omega_3$ follows from the assumed associativity of the four-fold products of the unitary operators $U(g)$ and in that case the infinitesimal generators fail to satisfy the Jacobi identity (\ref{jacobi1}).    

\begin{figure}
 \centering
\includegraphics[angle=0,width=6cm]{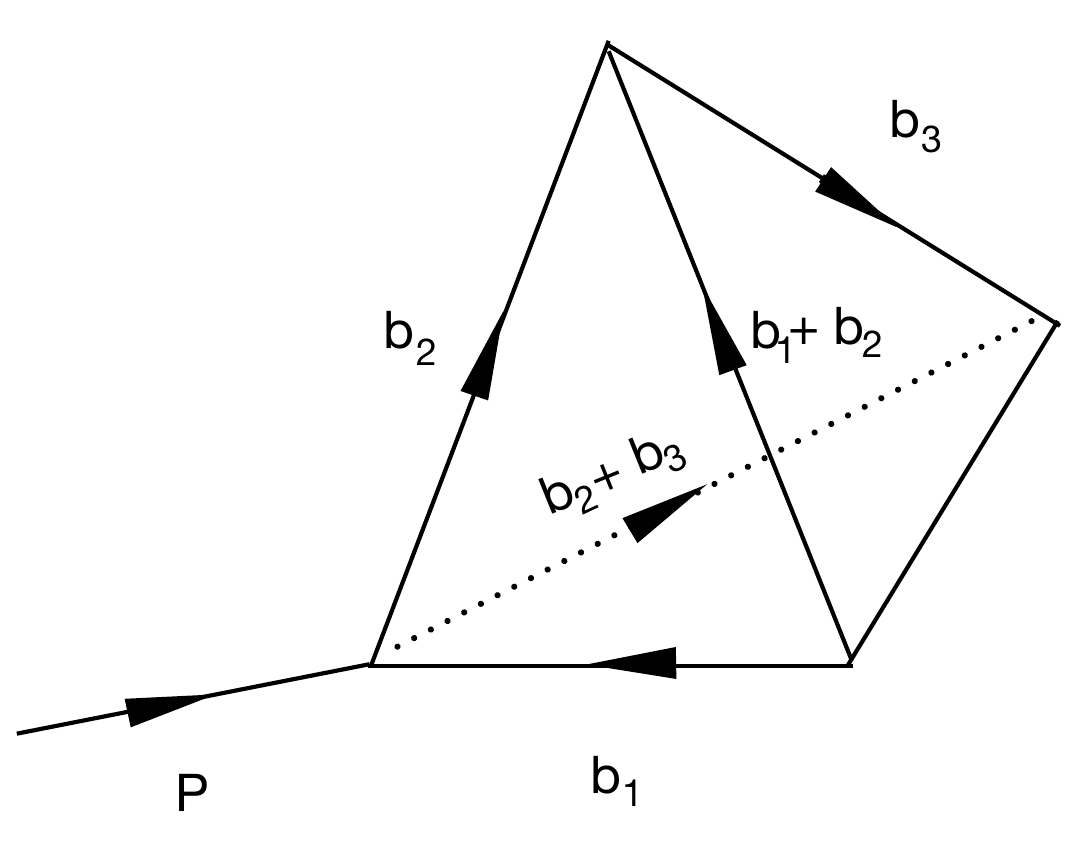}
\centering
\caption{  The expression $ \{(\vec{b}_2  \times  \vec{b}_3 )  +  (\vec{b}_1  \times ( \vec{b}_2 + \vec{b}_3 ))  -     (\vec{b}_1  \times  \vec{b}_2 )  -   ( (\vec{b}_1+\vec{b}_2)  \times  \vec{b}_3 )   \}   \cdot  \vec{P}  /P^3 $  is equal to the total flux of the momentum $\vec{P}/P^3$ through the surface of the tetrahedron $(\vec{b}_1, \vec{b}_2,\vec{b}_3)$. }
\label{fig1} 
\end{figure}

Let us evaluate  the products in (\ref{associ})  in order to obtain the explicit  expression for the phase factor $\Phi$
\be
U(\vec{b}_2) U(\vec{b}_3)  = e^{-{i\over \hbar}( \vec{b}_2+ \vec{b}_3) \hat{\vec{R}}    + {1\over 2 \hbar^2}  [i \vec{b}_2 \hat{\vec{R}}, ~ i \vec{b}_3 \hat{\vec{R}}  ]} = e^{ {i \lambda\over 2  \hbar}   (\vec{b}_2  \times  \vec{b}_3 ) \cdot {\vec{P} \over P^3} - {i \over \hbar}( \vec{b}_2+ \vec{b}_3) \hat{\vec{R}}  } = e^{i \Phi_{23}/\hbar}  U(\vec{b}_2 + \vec{b_3}),
\ee
where the phase $e^{i  \Phi_{23} /\hbar} $  is a flux $\Phi_{23} = \lambda \vec{S}_{23}   {\vec{P} \over P^3}$ of the momentum ${\vec{P} \over P^3} $ through the surface $\vec{S}_{23} $ of the triangle with momentum sites $\vec{b}_2$ and $\vec{b}_3$. We can now calculate the phase in the associativity relation  
\beqa
e^{{i \over \hbar}\Phi} = U(\vec{b}_1) ~( U(\vec{b}_2) U(\vec{b}_3) ) ~((U(\vec{b}_1) ( U(\vec{b}_2)) ~U(\vec{a}_3))^{-1} =\nn\\
\exp{\Big[ {i \lambda\over 2 \hbar }  \{ (\vec{b}_2  \times  \vec{b}_3 )  +  (\vec{b}_1  \times ( \vec{b}_2 + \vec{b}_3 ))  -     (\vec{b}_1  \times  \vec{b}_2 )  -   ( (\vec{b}_1+\vec{b}_2)  \times  \vec{b}_3 )   \}\cdot  \vec{P} / P^3  \Big]} 
\eeqa
and obtained that the phase factor in  (\ref{associ})  is the total flux of the momentum ${\vec{P} \over P^3} $ through the surface of tetrahedron constructed by the three momentum vectors  $(\vec{b}_1, \vec{b}_2,\vec{b}_3)$ (see Fig.\ref{fig1})
\be
\Phi = \lambda \oint {\vec{P} \over P^3} d \vec{S}= \lambda \int \Big(\vec{\nabla}\cdot {\vec{P} \over P^3} \Big)  ~d^3 \vec{P}=
\lambda \int  4 \pi \delta^{(3)}(\vec{P}) ~d^3 \vec{P} = 4 \pi \lambda
\ee
The quantum mechanical associativity will be restored if the phase $\Phi$ is proportional to $2 \pi n$ thus we will get\footnote{The non-associating quantities cannot be represented by well-defined linear operators, acting on a vector or Hilbert space, since by definition operators on vectors necessarily associate.
} 
\be\label{helicityqunatisation}
{\Phi \over \hbar } = {4 \pi \lambda\over \hbar }   = 2 \pi n, ~~~~~\lambda = {\hbar \over 2} n,~~~n = \pm 1, \pm 2, .....
\ee
{\it The quantum mechanical consistency requires the quantisation of the particles helicity.}  In the next section we will discuss the correspondence, a duality map,  between helicity quantisation (\ref{helicityqunatisation})  and the Dirac quantisation condition for the magnetic charge.  

\section{\it Correspondence with Dirac quantisation condition}

Consider a particle of mass $m$ and charge $e$ moving in a magnetic field of a monopole of charge $g_m$. The  Hamiltonian of a particle is
\be
H = {1\over 2 m} (\vec{p} -e \vec{A})^2,    
\ee
where $\vec{p}$ is  the canonical momentum  and the Heisenberg commutation relations are 
\beqa\label{partinmag}
~[x_n , x_m]=0,~~~~~~~~[x_n, p_m]= i \hbar \delta_{nm},~~~~~~~~ 
~[p_n, p_m]= i \hbar {e\over c} \epsilon_{nmk} H_k,
\eeqa
where the magnetic field of a monopole is 
\be
\vec{H} =  g_m  {\vec{r} \over r^3}.
\ee 
One can verify that the Jacobi identity of triple momentum commutator has the following form:
\be\label{partmagassoci}
~[[p_1,p_2],p_3] + [[p_2,p_3],p_1] +[[p_3,p_1],p_2] = 
\hbar^2 {e g_m\over  c} ~ \triangle_{ r} \big({1\over r}\Big)   = - 4 \pi \hbar^2 {e g_m\over  c} ~\delta^3(\vec{r})
\ee 
Let us consider the {\it coordinate translation operator}  \cite{Jackiw:1984rd} 
\be
U(\vec{a}) = e^{{i\over \hbar} \vec{a} \vec{p}}, ~~~~~U(\vec{a}) \psi(\vec{r})= \psi(\vec{r}+ \vec{a})
\ee 
and the corresponding associativity relation 
 \be\label{associ1}
U(\vec{a}_1) ~( U(\vec{a}_2) U(\vec{a}_3) ) = e^{i\Phi_m/\hbar} ~(U(\vec{a}_1) ( U(\vec{a}_2)) ~U(\vec{a}_3). 
 \ee
 One can calculate the phase in the associativity relation  
\beqa
e^{{i \over \hbar}\Phi_m} = U(\vec{a}_1) ~( U(\vec{a}_2) U(\vec{a}_3) ) ~((U(\vec{a}_1) ( U(\vec{a}_2)) ~U(\vec{a}_3))^{-1} =\nn\\
\exp{\Big[i {  e g_m  \over 2 \hbar c }  \{ (\vec{a}_2  \times  \vec{a}_3 )  +  (\vec{a}_1  \times ( \vec{a}_2 + \vec{a}_3 ))  -     (\vec{a}_1  \times  \vec{a}_2 )  -   ( (\vec{a}_1+\vec{a}_2)  \times  \vec{a}_3 )   \}\cdot  \vec{r} / r^3  \Big]} 
\eeqa
and showing  that the phase is the total flux of the vector field  ${\vec{r} \over r^3} $ through the surface of tetrahedron constructed by the three space-like vectors  $(\vec{a}_1, \vec{a}_2,\vec{a}_3)$  \cite{Jackiw:1984rd} 
\be
\Phi_m ={e g_m \over c} \oint {\vec{r} \over r^3} d \vec{S}= {e g_m \over c} \int \Big(\vec{\nabla}\cdot {\vec{r} \over r^3}\Big)  ~d^3 \vec{r}=
{e g_m \over c} \int  4 \pi \delta^{(3)}(\vec{r}) ~d^3 \vec{r} = 4 \pi {e g_m \over c}
\ee
The quantum mechanical associativity will be restored if the phase in proportional to $2 \pi n$ 
\be\label{Diracquant}
{\Phi_m \over \hbar } = 4 \pi  {e g_m \over \hbar c}   = 2 \pi n, ~~~~~{e g_m \over c}  = {\hbar \over 2} n,~~~n = \pm 1, \pm 2, .....
\ee
thus recovering the Dirac quantisation condition. The removal of the 3-cocycle, which is necessary for conventional quantum mechanics on the operators on Hilbert space, limits magnetic sources to quantised Dirac monopoles\footnote{ In that respect it is interesting to note that it was argued by Saha \cite{Saha:1949np} that a charge quantisation condition can also be derived from quantisation of angular momentum of a particle moving in a monopole background.  It is noted in  \cite{Goddard:1977da}, "This argument is plausible but too vague to be convincing as it stands. For example, since no fermions are involved it might seem more reasonable to have the components of $J$ quantised in integral, rather than half-integral, multiples of $\hbar$."  (see also \cite{Deguchi:2005qd}).}. Other magnetic sources lead to non-associative algebras\footnote{The non-associative algebras where considered in \cite{Witten:1985cc, Savvidy:2002gs,Blumenhagen:2011ph,Bojowald:2018qqa,Szabo:2017yxd,Szabo:2019hhg, Mylonas:2013jha, Schupp:2023jda}.}.

The Dirac quantisation condition, represented by the relations (\ref{partinmag}), (\ref{partmagassoci}) and (\ref{Diracquant}),  is in a well defined correspondence (duality) with the relations (\ref{masslesscood1}), (\ref{jacobi1}) and (\ref{helicityqunatisation}) if one consider a map 
\be\label{map}
\hat{\vec{R}} \longleftrightarrow  \vec{p},~~~~\vec{P}  \longleftrightarrow - \vec{r}, ~~~~~~~~\lambda  \longleftrightarrow {e g_m\over  c}. 
\ee
{\it In summary, we have demonstrated that there is a well defined correspondence (duality) (\ref{map}) between helicity quantisation condition of massless relativistic particles (\ref{helicityqunatisation})  and the Dirac quantisation condition of the magnetic charge(\ref{Diracquant}), which is based on the analyses of 3-cocycles in the corresponding associativity relations (\ref{associ}) and (\ref{associ1})}.

\section{\it Photons and Gravitons }

The non-commutativity of the space coordinates of the massless particles (\ref{masslesscood1}) makes it more natural to represent the wave function of massless particles in a momentum representation (\ref{uniop}) rather than in the coordinate representation because  here the momentum components  are commuting (\ref{heizenberg3}) and have the standard commutation relations with space coordinates. 

This phenomenon is noticeable in quantum electrodynamics when considering the wave function of the massless photons.  In referring to the photon wave function, it was emphasised in  \cite{Berestetskii, Landau1930} that it cannot be regarded as the probability amplitude of the spatial localisation of the photon, in contrast to the fundamental significance of the wave function in non-relativistic quantum mechanics. 

Mathematically, this is shown by the impossibility of constructing from the coordinate wave function any quantity that has formal properties of a probability density \cite{Berestetskii}.  The wave function in the momentum representation has a more profound physical significance than that in the coordinate representation: it enables to calculate the probabilities $w({\bf k},  {\bf e}_{\alpha})$ of various values of the momentum and polarisation $ {\bf e}_{\alpha}$ of a photon in a specified state \cite{Berestetskii}. It is given by the square of the modulus of the corresponding coefficient in the expansion of the wave functions in states with given ${\bf k}$ and ${\bf e}_{\alpha}$ \cite{Berestetskii}\footnote{The $w$ is invariant under  gauge  transformation ${\bf A}({\bf k})  \rightarrow {\bf A}({\bf k})  + \alpha {\bf k}$, as far as ${\bf e}_{\alpha}  \cdot  {\bf k}   =0$. }
\be
w({\bf k},  {\bf e}_{\alpha}) \propto \vert {\bf e}_{\alpha} \cdot {\bf A}({\bf k}) \vert^2 .
\ee
It seems that an additional conclusion can be  made also with respect to the massless gravitons. The natural relevance of  gravitons to the geometry of the spacetime physics and non-commutativity of their  space coordinates may lead to a deeper understanding of the quantum gravity at short and large distances.  In particular, we suggested that there should be a minimal space cell volume (\ref{quantvol}), similar to the minimal phase space cell of Heisenberg, $2 \pi \hbar$.  The space-time quantisation was considered in the earlier publications  in \cite{Doplicher:1994zv, Doplicher:1994tu, Yoneya:2000bt, Mylonas:2013jha}\footnote{I would like to thank Richard Szabo for pointing me to these references.}. 

The non-commutativity of space coordinates was derived in the string theory framework  in series of publications by Lust and collaborators  in  \cite{Lust:2010iy, Condeescu:2012sp,  Andriot:2012vb, Bakas:2013jwa, Hassler:2013wsa}. The commutator relations that were derived in different string backgrounds have in general  the following form:
\be
~[x_n, x_m] = i \hbar R_{nml} p^l, 
\ee
where $(x_n,p_m)$ are zero modes of string position and its conjugate momentum in non-geometric background fluxes   $R_{nml}$.  

When one compares this relation with (\ref{masslesscood}), one can notice a great similarity and difference as well: both relations involved the momentum operators on the right-hand side of the commutator, while behaviour of the commutator at large and small momenta is different. The commutator (\ref{masslesscood}) decreases in ultraviolet region and is singular in infrared leading to the natural quantisation of particle's helicity (\ref{helicityqunatisation}).  It seems that these differences have a deep meaning and have to be understood better.

\section{\it High-spin extension of Poincar\'e algebra}

  Let us also consider the extension of the high-spin Poincar\'e algebra introduced earlier in \cite{Savvidy:2008zy, Savvidy:2010vb, Savvidy:2010kw}. The algebra (\ref{extensionofpoincarealgebra}) naturally appeared in the high-spin extension of the Yang Mills theory suggested in \cite{Savvidy:2005fi, Savvidy:2015jgv,Savvidy:2006yk,Savvidy:2006yj}.   The  non-Abelian tensor gauge fields  were defined as rank-$(s+1)$ tensors
$$
A^{a}_{\mu\lambda_1 ... \lambda_{s}}(k),~~~~~s=0,1,2,...,
$$
and were totally symmetric with respect to the indices $  \lambda_1 ... \lambda_{s}  $ and had no symmetries with
respect to the first index  $\mu$. The index $a$ numerates the generators $L^a$ of the Lie algebra  of a compact Lie group G. These tensor fields appear in the expansion of the gauge field $\CA_{\mu}(k,e)$ over the unit  polarisation vector
$e_{\lambda}$  \cite{Savvidy:2015jgv} :
\be\label{gaugefield}
{\cal A}_{\mu}(k,e)=\sum_{s=0}^{\infty} {1\over s!} ~A^{a}_{\mu\lambda_{1}...
\lambda_{s}}(k)~L_{a}^{\lambda_1 ... \lambda_{s}} , 
\ee
where $L_{a}^{\lambda_1 ... \lambda_{s}} =e^{\lambda_1}...e^{\lambda_s} \otimes L_a $  are the "gauge generators" of the following current algebra  \cite{Savvidy:2008zy, Savvidy:2010vb, Savvidy:2010kw}:
\be\label{curralg}
[L_{a}^{\lambda_1 ... \lambda_{i}}, L_{b}^{\lambda_{i+1} ... \lambda_{s}}]=if_{abc}~
L_{c}^{\lambda_1 ... \lambda_{s} } ,~~~~~s=0,1,2,...
\ee
This current  algebra has infinitely many "gauge generators" $L_{a}^{\lambda_1 ... \lambda_{s}} $. They are gauge generators because they carry the isospin and Lorentz indices. The  generators $L_{a}^{\lambda_1 ... \lambda_{s}} $ are symmetric spacetime tensors, and  the full algebra  includes the Poincar\'e generators $P_{\mu},~M_{\mu\nu}$.  The algebra $L_G(\CP)$  has the following form \cite{Savvidy:2008zy, Savvidy:2010vb, Savvidy:2010kw} :
\beqa\label{extensionofpoincarealgebra}
~&&[P^{\mu},~P^{\nu}]=0,\nn\\
~&&[P^{\lambda} ,~M^{\mu\nu}] = i \hbar (\eta^{\lambda \nu}~P^{\mu}
- \eta^{\lambda \mu }~P^{\nu}) ,\nn\\
~&&[M^{\lambda \rho},~M^{\mu \nu} ] = i \hbar (\eta^{\mu \rho}~M^{\nu \lambda}
-\eta^{\mu \lambda}~M^{\nu \rho} +
\eta^{\nu \lambda}~M^{\mu \rho}  -
\eta^{\nu \rho}~M^{\mu \lambda} ),\nonumber\\
~&&[P^{\mu},~L_{a}^{\lambda_1 ... \lambda_{s}}]=0, \nn\\
~&&[L_{a}^{\lambda_1 ... \lambda_{s}},~M^{\mu \nu} ] = i \hbar(
\eta^{\lambda_1\nu } L_{a}^{\mu \lambda_2... \lambda_{s}}
-\eta^{\lambda_1\mu} L_{a}^{\nu\lambda_2... \lambda_{s}}
+...+
\eta^{\lambda_s\nu } L_{a}^{\lambda_1... \lambda_{s-1}\mu } -
\eta^{\lambda_s\mu } L_{a}^{\lambda_1... \lambda_{s-1}\nu } ),\nonumber\\
~&&[L_{a}^{\lambda_1 ... \lambda_{i}}, L_{b}^{\lambda_{i+1} ... \lambda_{s}}]=if_{abc}~
L_{c}^{\lambda_1 ... \lambda_{s} } ,     ~~~(\mu,\nu,\rho,\lambda=0,1,2,3; ~~~~~s=0,1,2,... )
\eeqa
It is an extension of the Poincar\'e algebra by "gauge generators" $L_{a}^{\lambda_1 ... \lambda_{s}} $, which are the elements of the current algebra (\ref{curralg}).   The  algebra (\ref{extensionofpoincarealgebra}) can be extended to a supersymmetric case  as well \cite{Antoniadis:2011re}.  The supersymmetric generalisations of high-spin de Sitter and conformal groups  were also suggested in \cite{Antoniadis:2011re}.   The algebra $L_G(\CP)$ has a representation in terms of differential operators of the
following general form:
\beqa\label{represofextenpoincarealgebra}
~&& P^{\mu} = k^{\mu} ,\nn\\
~&& M^{\mu\nu} = i\hbar (k^{\nu}~ {\partial\over \partial k_{\mu}}
- k^{\mu }~ {\partial \over \partial k_{\nu}}) + i \hbar (e^{\nu}~ {\partial\over \partial e_{\mu}}
- e^{\mu }~ {\partial \over \partial e_{\nu}}),\nn\\
~&& L_{a}^{\lambda_1 ... \lambda_{s}} =e^{\lambda_1}...e^{\lambda_s} \otimes L_a,
\eeqa
where $e^{\lambda} $ is a translationally invariant space-like unit vector.
The vector space of a representation  is parameterised
by the momentum $k^{\mu}$ and  vector variables $e^{\lambda}$:
$
\Psi(k^{\mu}, e^{\lambda} )~.
$
Irreducible representations can be obtained from (\ref{represofextenpoincarealgebra})
by   imposing  invariant constraints on the
vector space  of the following form:
\be\label{constraint}
k^2=0,~~~k^{\mu} e_{\mu}=0,~~~e^2=1~.
\ee
These equations have a unique solution  
$
e^{\mu}= \chi k^{\mu} + e^{\mu}_{1}\cos\varphi +e^{\mu}_{2}\sin\varphi,
$
where $e^{\mu}_{1}=(0,1,0,0),~ e^{\mu}_{2}=(0,0,1,0)$ when $k^{\mu}=\omega(1,0,0,1)$.
The  $\chi$ and  $\varphi$ remain as independent variables
on the cylinder $ \varphi \in S^1, \chi \in R^1 $. The invariant subspace of functions  now reduces to
the following form:
$
\Psi(k^{\mu}, e^{\nu} )~\delta(k^2)~\delta(k\cdot e)~\delta(e^2 -1)
= \Phi(k^{\mu}, \varphi, \chi) .
$
The generators
$L_{a}^{ \lambda_1 ... \lambda_{s}} $  take the following form:
\be\label{trasversalgenera}
L_{a}^{\bot~ \lambda_1 ... \lambda_{s}}= \prod^{s}_{n=1} ( \chi k^{\lambda_n} + e^{\lambda_n}_{1}\cos\varphi
+e^{\lambda_n}_{2}\sin\varphi) \otimes L_a.
\ee
This is a purely transversal representation because of (\ref{constraint}):
\be\label{transversality}
k_{\lambda_1}L_{a}^{\bot \lambda_1 ... \lambda_{s}}=0,~~~~s=1,2,...
\ee
The generators
$L_{a}^{\bot~ \lambda_1 ... \lambda_{s}}$  carry helicities in the following range:
\be\label{trasversalgenera1}
\lambda=(s,s-2,......, -s+2, -s),
\ee
in total $s+1$ states.  This can be deduced from the explicit representation (\ref{trasversalgenera}) by using
helicity polarisation vectors $e^{\lambda}_{\pm}= (e^{\lambda}_1 \mp i e^{\lambda}_2)/2$:
\be\label{trasversalgenera11}
L_{a}^{\bot~ \lambda_1 ... \lambda_{s}}= \prod^{s}_{n=1} ( e^{i \varphi} e^{\lambda_n}_{+}
+e^{-i \varphi} e^{\lambda_n}_{-}   + \chi k^{\lambda_n} ) \otimes L_a.
\ee
After performing the multiplication
in (\ref{trasversalgenera11}) and collecting the terms of a given power
of momentum we will get the following expression:
\beqa\label{trasversalgenera2}
L_{a}^{\bot~ \mu_1 ... \mu_{s}}= \prod^{s}_{n=1} (e^{i \varphi} e^{\mu_n}_{+}
+e^{-i \varphi} e^{\mu_n}_{-}) \otimes L_a +~~~~~~~~~~~~~~~~~~~~~~~~~~~~~~~~~~~~~~~~~~~~~~ \\
+\sum_{1} \chi k^{\mu_1} \prod^{s}_{n=2} (e^{i \varphi} e^{\mu_n}_{+}
+e^{-i \varphi} e^{\mu_n}_{-}) \otimes L_a +...+\chi
k^{\mu_1}... \chi k^{\mu_s} \otimes  L_{a} ,\nn
\eeqa
where the first term
$\prod^{s}_{n=1} (e^{i \varphi} e^{\mu_n}_{+}
+e^{-i \varphi} e^{\mu_n}_{-})$ represents the {\it helicity generators }
$(L^{+\cdot\cdot\cdot+}_{a},...,L^{-\cdot\cdot\cdot-}_{a})$, while their
helicity spectrum is described by the formula  (\ref{trasversalgenera1}).
 The transversal representation  $L_{a}^{\bot \lambda_1 ... \lambda_{s}}$    defines the helicities  of the gauge field $\CA_{\mu}(k,e)$ in (\ref{gaugefield}). 
By substituting the transversal representation (\ref{trasversalgenera2}) of the
generators $L_{a}^{\bot \lambda_1 ... \lambda_{s}}$ into the
expansion (\ref{gaugefield}) and collecting the terms in front of the helicity generators
$(L^{+\cdot\cdot\cdot+}_{a},...,L^{-\cdot\cdot\cdot-}_{a})$  we will get
\beqa\label{gaugefield1}
{\cal A}_{\mu}(k,e)&=&  \sum_{s=0}^{\infty} {1\over s!} ~
 (\tilde{A}^{a}_{\mu \lambda_1 ... \lambda_{s}}~e^{\lambda_1}_{+}...e^{\lambda_s}_{+} \otimes L_{a} +...+
\tilde{A}^{a}_{\mu \lambda_1 ... \lambda_{s}} e^{\lambda_1}_{-}...e^{\lambda_s}_{-} \otimes L_{a} )     \nn\\
&=& \sum_{s=0}^{\infty} {1\over s!} ~
 (\tilde{A}^{a}_{\mu +\cdot\cdot\cdot+}~\otimes L^{+\cdot\cdot\cdot+}_{a} +...+
\tilde{A}^{a}_{\mu -\cdot\cdot\cdot-}~\otimes L^{-\cdot\cdot\cdot-}_{a} ),
\eeqa
where s is the number of negative indices.
This formula represents the projection  $\tilde{A}^{a}_{\mu \lambda_1 ... \lambda_{s}}$
 of the components  of
the non-Abelian tensor gauge field  $A^{a}_{\mu\lambda_1 ... \lambda_{s}} $ into the plane
transversal to the momentum.  The projection contains only positive-definite space-like
components of the helicities \cite{Savvidy:2010vb,Savvidy:2010kw,Savvidy:2015jgv}:
\be\label{tensorspectrum}
\lambda~~=~~~\pm (s+1),~~ \begin{array}{c} \pm (s-1)\\ \pm (s-1) \end{array},~~
\begin{array}{c} \pm (s-3)\\ \pm (s-3) \end{array},~~....,
\ee
where the lower helicity states have double degeneracy.  

In summary, we have seen that the spectrum of the high-spin extension of Poincar\'e algebra is massless and together with the unbounded nature of the $\lambda$ spectrum (\ref{tensorspectrum}), ranging over all integers, indicates that physically accessible states would exist, for which $(\triangle \hat{\textbf{R}}  )^2  \equiv \triangle \hat{R_1}^2 +\triangle \hat{R_2}^2 +\triangle \hat{R_3}^2 $  is arbitrarily large due to the uncertainty  relation (\ref{uncertanty}).

\section{\it Acknowledgments}
 
  I would like to thank Richard Szabo and Dieter Lust for the delightful communication and for providing me references  on  the earlier publications on the space-time quantisation and Konstantin Savvidy for stimulating discussions and helping to clarify the uncertainty relations (\ref{volumeuncer}) and (\ref{volumeuncer1}).
   
\section{\it Appendix}
The commutator of the new non-commutative coordinates with $P^0$ have the following form 
\beqa
&~[\hat{R}_n, P^0 ]=  i \hbar {P_n\over P^0}
\eeqa
and allows to calculate the following commutators
\beqa
&~[P^0 \hat{R}_n, P^0 \hat{R}_m]= -i \hbar  \lambda \epsilon_{nmk} {P_k \over P^0} + i \hbar ( \hat{R}_m P_n -\hat{R}_n P_m ) \\
&~ [ \hat{R}_n P^0 ,  \hat{R}_m  P^0]=  -i \hbar  \lambda \epsilon_{nmk} {P_k \over P^0} + i \hbar (P_n   \hat{R}_m - P_m \hat{R}_n  ) \nn\\
& ~[P^0 \hat{R}_n, \hat{R}_m P^0 ]=  -i \hbar  \lambda   \epsilon_{nmk} {P_k \over P^0}   + i \hbar (\hat{R}_m P_n  - P_m \hat{R}_n )  - \hbar^2 {P_nP_m \over( P^0)^2} \nn\\
&~ [\hat{R}_n P^0  , P^0  \hat{R}_m  ]= -i \hbar   \lambda  \epsilon_{nmk} {P_k \over P^0}   + i \hbar ( P_n \hat{R}_m    - \hat{R}_n P_m )  + \hbar^2 {P_nP_m \over( P^0)^2}  \nn
 \eeqa

\bibliographystyle{elsarticle-num}
\bibliography{poincare}

\vfill
\end{document}